# Multimedia-based Medicinal Plants Sustainability Management System

Zacchaeus Omogbadegun[1,§], Charles Uwadia[2], Charles Ayo[3], Victor Mbarika[4], Nicholas Omoregbe[5], Efe Otofia[6], Frank Chieze[7]

[1]Computer and Information Sciences Department, College of Science and Technology, Covenant University,
Ota, Ogun State, Nigeria

[2]Department of Computer Sciences, University of Lagos,
Lagos, Nigeria

[3]Computer and Information Sciences Department, College of Science and Technology, Covenant University,
Ota, Ogun State, Nigeria

[4]Southern University and A&M College,
Baton Rouge, LA 70813, USA

[5]Computer and Information Sciences Department, College of Science and Technology, Covenant University,
Ota, Ogun State, Nigeria

[6]Computer and Information Sciences Department, College of Science and Technology, Covenant University,
Ota, Ogun State, Nigeria

[7]Electrical and Information Engineering Department, College of Science and Technology, Covenant University,
Ota, Ogun State, Nigeria

[§]Corresponding author

**Abstract**

Medicinal plants are increasingly recognized worldwide as an alternative source of efficacious and inexpensive medications to synthetic chemo-therapeutic compound. Rapid declining wild stocks of medicinal plants accompanied by adulteration and species substitutions reduce their efficacy, quality and safety. Consequently, the low accessibility to and non-affordability of orthodox medicine costs by rural dwellers to be healthy and economically productive further threaten their life expectancy. Finding comprehensive information on medicinal plants of conservation concern at a global level has been difficult. This has created a gap between computing technologies' promises and expectations in the healing process under complementary and alternative medicine. This paper presents the design and implementation of a Multimedia-based Medicinal Plants Sustainability Management System addressing these concerns. Medicinal plants' details for designing the system were collected through semi-structured interviews and databases. Unified Modelling Language, Microsoft-Visual-Studio.Net, C#3.0, Microsoft-Jet-Engine4.0, MySQL, Loquendo Multilingual Text-to-Speech Software, YouTube, and VLC Media Player were used.

*Keywords*: Complementary and Alternative Medicine, conservation, extinction, medicinal plant, multimedia, phytoconstituents, rural dwellers

## 1. Introduction

Plants and animals hold medicinal, agricultural, ecological, commercial and aesthetic/recreational value. Some plants of medicinal value: *Anacardium occidentale (*Cashew nut), *Azadirachta indica (*Neem), *Allium sativum* L.(Garlic), and *Zingiber officinale* Roscoe (Common Ginger), are shown in **Fig. 1**.

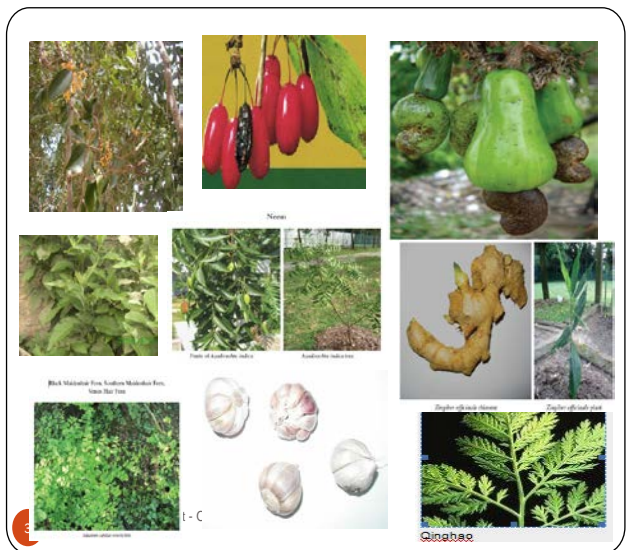

Fig. 1 Some plants of medicinal value

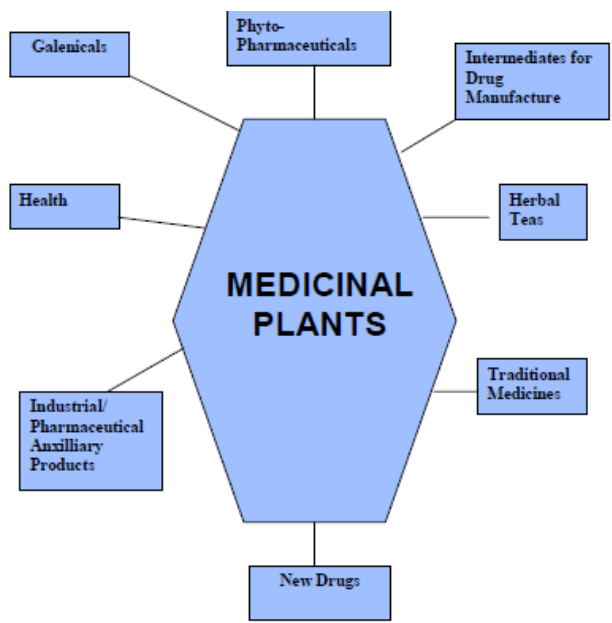

Fig. 2 Industrial Uses of Medicinal Plants [5]

Medicinal plants have become the most important source of life-saving drugs for the majority of the world's population. Medicinal plants harvested from the wild remain of immense importance for the well-being of millions of people around the world. Over 70,000 plant species are thought to be medicinal **[1]**.

Medicinal plants are considered a source of various alkaloids and other chemical substances essential for mankind. Over 80% of the US public uses nonconventional practices and complementary medicines adjunctive to conventional medical care. According to the World Health Organization, over 80% of the people in developing countries depend upon traditional medicine for their primary health care **[2]**.

Africa has been and continues to be a significant source of medicinal and aromatic plants and botanicals to the world's food, drug, herb and dietary supplement market. About 50% of drugs used in modern medicine are of plant origin. About 80% of Africa's population rely on medicinal plants for their health needs confirming that medicinal plant preparations have been identified as alternative remedies for several diseases **[3]**. The active principles of many plant species are isolated for direct use as drugs, lead compounds or pharmacological agents **[4]**.

The medicines for internal use prepared in the traditional manner involve simple methods such as hot- or cold-water extraction, extraction of juice after crushing, powdering of dried material, formulation of powder into pastes via such a vehicle as water, oil or honey, and even fermentation after adding sugar source. The range of products that could be obtained from medicinal plants is given in **Fig**.**2 [5]**.

Low accessibility to and affordability of orthodox medicine by rural dwellers and their need to keep healthy to be economically productive have led to their dependence on medicinal plants to remedy afflictions **[6]**. In Nigeria herbal practices, the practitioners claim that plant parts possess various phytochemicals which exhibit diverse pharmacological and biological responses and diversities. Nigeria is a country stepped in the use of and belief in traditional medicines in which plants play a major role **[7]**.

**1.1 Statement of the Problem**

Emerging new infectious, chronic and drug-resistant diseases have prompted scientists to look towards medicinal plants as agents for treatment and prevention.

Medicinal plants' species are threatened by habitat loss, climate change, and species-specific, multipurpose over-harvesting and logging leading to potential extinction of useful medicinal plants in the continent.

Securing supplies of quality products before the over-harvesting of wild stocks depletes the resource constitutes a concern.

Declining wild stocks of medicinal plants are accompanied by adulteration and species substitutions, which in turn reduce efficacy, quality and safety.

Imperceptibly, these medicinal plants' sustainability remains in jeopardy creating a gap between computing technologies' promises and expectations in the healing process under complementary and alternative medicine (CAM).

Difficulty encountered in finding comprehensive information on medicinal plants at a global level to promote scientific research towards obtaining clues and discovery of potential lead compounds and novel therapeutics.

Improvement in the quality of life of the rural poor; development of traditional medicines and reduction of the overexploitation of plants are inevitably desirable.

### 1.2 Research Questions

The following Research Questions have been raised for attention in this paper: (1) How can empirical knowledge of medicinal plant uses, often held by an older generation of healers in remote areas, be accumulated, stored and transmitted to next generations without compromising their intellectual property rights? (2) How can consumers be protected from false information or the use of products with negative side-effects? (3) How can sustainable wild sourcing be implemented – or the medicinal plants be 'domesticated' – to secure supplies of quality products before the over-harvesting of wild stocks depletes the resource?

### 1.3 Objectives of the Research

Our objectives included (1) providing a platform for a multidisciplinary team of scholars and healthcare services providers / CAM practitioners for information exchange in African healing process seamlessly;
(2) designing a framework for conserving, protecting and propagating medicinal plants, animals and cultural sites across the African continent; and
(3) implementing a system which would facilitate efficient knowledge discovery on medicinal plants with voice/video features on a multimedia platform.

### 2. Patients' Healthcare Requirements

Patients' requirements for healthcare include treatment and care that work, good relationship with practitioner, provision of information, and remaining in control of treatment. Complementary and alternative medicine continues to attract patronage due to patients' dissatisfaction with conventional health care, a desire for greater control over one's health, and a desire for cultural and philosophical congruence with personal beliefs about health and illness **[8].**

Nigeria is rich in biodiversity. The country is endowed with a variety of plant and animal species. As reflected in **Table 1,** there are about 7, 895 plant species identified in 338 families and 2, 215 genera **[9].**

Table 1 Plant species in Nigeria [9]

| Groups Of Plants | Families | Genera | Species |
|---|---|---|---|
| Algae | 67 | 281 | 1335 |
| Lichens | - | 14 | 17 |
| Fungi (Mushrooms) | 26 | 60` | 134 |
| Mosses | - | 13 | 16 |
| Liverworts | - | 16 | 6 |
| Pteridophytes | 27 | 64 | 165 |
| Gymnosperms | 2 | 3 | 5 |
| Chlamydosperms | 2 | 2 | 6 |
| Monocotyledons | 42 | 376 | 1575 |
| Dicotyledons | 172 | 1396 | 4636 |
| Total | 338 | 2215 | 7895 |

The expanding trade in Medicinal Plants has serious implications on the survival of several plant species, with many under serious threat to become extinct. Recently however, attention is turning back to natural products as drug sources, since they have been so successful in the past. Modern medicine depends on biological materials as an incomparable source of molecular diversity. Against this backdrop, almost half the world's plant species may be threatened with extinction; cures as yet undiscovered may exist in plants as yet un-described - and which may never be described. Promising drug sources are also found in the sea - sponges, sea squirts and algae for example, are all sources of drugs undergoing clinical studies. Plants are the structural anchors of the ecosystems in which these organisms live. The rapid loss of plant life has far-reaching consequences, and their loss will adversely affect future drug discovery **[1].**

### 2.1 Medicinal Plants Endangered

Due to irresponsible human acts of mass destruction of forests worldwide, we are losing flora at an alarming rate. Unless we act immediately to preserve the medicinal plants, plants with nutraceutical values, future generations will loose tremendous health and wellness benefits from nutraceutical herbs that we are enjoying now **[10].**

Despite the long tradition of usage of medicinal plants, their proven efficacy, and lack of affordable alternatives, the continued availability of many of these plants is in jeopardy as their species are threatened by habitat loss, climate change, multipurpose over-harvesting, and logging.

Many medicinal plants are being destroyed at an unprecedented rate and are threatened with extinction. The destruction of plant species is occurring at a rate unmatched in geological history. Current extinction rates are at least 100 to 1,000 times higher than natural background rates, with a quarter of the world's coniferous trees in jeopardy, and as many as 15,000 medicinal plants threatened **[11]**.

In USA, approximately 250,000 species of flowering plants, it is estimated that some 60,000 of these may become extinct by the year 2050, and more than 19,000 species of plants are considered to be threatened or endangered from around the world. More than 2000 species of plants native to the United States are threatened or endangered, with as many as 700 species becoming extinct in the next 10 years [12].

In Pakistan, it was observed that 49 medicinal plants species belonging to 32 different families were sold in local markets and thus playing a role in uplifting the socioeconomic conditions of the area. It was observed that out of these 49 medicinal plants, 24 plant species are threatened (9 Endangered, 7 Vulnerable and 8 Rare) due to excessive collection from the wild. These plants are also used locally for curing different ailments. In most cases, the market availability status of these medicinal plants have increased, showing an increased inclination of local people towards medicinal plants collection and increased dependency of local population on medicinal plants trade. A brief set of information about these plants is given in **Table 2 [13].**

**Table 2 Folk medicinal uses, market availability status, conservation status of some important medicinal plants of Swat, Pakistan [13] (Extracts)**

| Plant Material | Family | Part Used | MS | CS |
|---|---|---|---|---|
| *Acorus calamus* L. | Araceae | Whole plant | P | E |
| *Berberis vulgaris* Linn | Berberidaceae | Whole plant | P | E |
| *Dioscorea deltoidea* Wall. | Dioscoreaceae | Tubers | D | E |
| *Polygonatum verticillatum* All. | Liliaceae | Rhizome | P | E |
| Paeoniaceae *Paeonia emodi* Wall. ex Hk.f. | Paeoniaceae | Rhizome/ seeds | P | E |
| *Podophyllum hexandrum* Royle | Podophyllaceae | Rhizome | P | E |
| *Bistorta amplexicaulis* (D.Don) Greene | Polygonaceae | Rhizome | P | E |
| *Bergenia ciliate* (Haw) Sternb. | Saxifragaceae | Leaves Rhizome | I | E |
| *Valeriana jatamansi* Jones | Valerianaceae | Rhizome | D | E |
| *Adiantum* capillus-veneris L. | Adiantaceae | Fronds | I | V |
| *Pistacia integerrima* Stew.ex Brand | Anacardiaceae | Leaves | I | V |
| *Berberis lyceum* Royle | Berberidaceae | Whole plant | I | V |
| *Ephedra gerardiana* Wall. ex Stapf | Ephedraceae | Fruit/ leaves | I | V |
| *Colchicum luteum* Baker. | Liliaceae | Rhizome/ seeds | I | V |

**Legend: MS = Market Status [ D = Decreased, I = Increased, P = Persistent], CS = Conservation Status [E = Endangered, V = Vulnerable]**

The downturn in the Nigerian economy and inflationary trend has led to the excessive harvesting of non-timber forest products for various uses. Some of these species are now threatened. Examples as reflected in **Table 3** are H*ymenocardia acida, Kigelia africana, and Cassia nigricans* **[9].**

**Table 3 Threatened Biodiversity Species in Nigeria [9]**

| SPECIES | MAIN USES | STATUS |
|---|---|---|
| **PLANTS** | | |
| *Milicea excelsia* | Timber | Endangered |
| *Diospyros elliotii* | Carving | Endangered |
| *Triplochiduiton scleroxylon* | Timber | Endangered |
| *Mansoiea altissinia* | Timber | Endangered |
| *Masilania accuminata* | Chewing stick | Endangered |
| *Garcina manni* | Chewing stick | Endangered |
| *Oucunbaca aubrevillei* | Trado-medical | Almost Extinct |
| *Erythrina senegalensis* | Medicine | Endangered |
| *Cassia nigricans* | Medicine | Endangered |
| *Nigella sativa* | Medicine | Endangered |
| *Hymenocardia acida* | General | Endangered |
| *Kigelia africana* | General | Endangered |

*The Red Data Book of India* has 427 entries of endangered species of which 28 are considered extinct, 124 endangered, 81 vulnerable, 100 rare and 34 insufficiently known species. In West Africa, Nigeria has the highest number of threatened species (119), followed by Ghana (115), Côte d'Ivore (101), Liberia (46) and Sierra Leone (43) **[14].**

A total of 54 different tree species (24 families) were identified in Ala, 41 species (21 families) in Omo and 55 species (20 families) in Shasha Forest Reserves of Nigeria. The most prevalent species in the ecosystem was *Strombosia pustulata*, while the family Leguminosae had the highest number of species. 84% of the species are regarded as rare or threatened with extinction while 16% were relatively abundant. Wanton removal of plant and animal species through over-harvesting activities is very inimical to biological conservation and has led to loss of biodiversity and extinction especially of many natural species with narrow range **[15]**. The status of medicinal plants under regular trade in the rainforests is presented in **Table 4**.

**Table 4 Respondent Opinions on the Status of Traded Medicinal Plants in Community Forests [15] (Extracts)**

| Medicinal plants | Opinion of respondents (%) | | | | |
|---|---|---|---|---|---|
| | Engd | Thrtd | Rar | Comn | % |
| *Alchornea cordifolia* | 44 | 24 | 20 | 12 | 100 |
| *Ananthus montanus* | 32 | 54 | 10 | 4 | 100 |
| *Bridelia ferruginea* | 30 | 48 | 22 | 24 | 100 |
| *Callichilia barteri* | 42 | 22 | 20 | 16 | 100 |
| *Canarium schweinfurthii* | 32 | 28 | 24 | 16 | 100 |
| *Cissus aralioides* | 34 | 48 | 10 | 8 | 100 |
| *Cocholospermum planchonni* | 44 | 26 | 16 | 14 | 100 |
| *Combretum smeathmanii* | 48 | 23 | 24 | 2 | 100 |
| *Enantia chloratha* | 44 | 30 | 14 | 12 | 100 |
| *Ocimum gratissimum* | 46 | 24 | 20 | 10 | 100 |
| *Rauvolfia vomitoria* | 24 | 64 | 8 | 4 | 100 |
| *Rauvolfia vomitoria* | 22 | 60 | 14 | 4 | 100 |
| *Rothmannia hispida* | 46 | 38 | 10 | 6 | 100 |
| *Sanseuieria guineense* | 24 | 60 | 14 | 2 | 100 |
| *Struchium spargonophora* | 32 | 48 | 22 | 18 | 100 |
| *Thorningia sanguinea* | 24 | 42 | 20 | 14 | 100 |
| *Uraria picta* | 44 | 22 | 14 | 20 | 100 |
| *Zingiber officinale* | 56 | 24 | 10 | 10 | 100 |

**Legend: Engd = Endangered, Thrtd = Threatened, Rar = Rare, Comn = Common**

The data were based on the opinions of stakeholders (respondents) directly involved in harvesting medicinal plants for the market. The data in **Table 4** revealed that 45 medicinal plants were traded on regular basis in the rainforest of Nigeria. Out of these, 8 medicinal plants were endangered, 12 species were rare and 8 species were threatened, while 17 were common or abundant in natural forests **[16].**

Plants and animals are responsible for a variety of useful medications. In fact, about forty percent of all prescriptions written today are composed from the natural compounds of different species. These species not only save lives, but they contribute to a prospering pharmaceutical industry worth over $40 billion annually. Unfortunately, only 5% of known plant species have been screened for their medicinal values, although we continue to lose up to 100 species daily. The Pacific yew, a slow-growing tree found in the ancient forests of the Pacific Northwest, was historically considered a "trash" tree (it was burned after clearcutting). However, a substance in its bark taxol was recently identified as one of the most promising treatments for ovarian and breast cancer. Additionally, more than 3 million American heart disease sufferers would perish within 72 hours of a heart attack without digitalis, a drug derived from the purple foxglove **[17].**

All over the world, much importance is being given to the documentation of knowledge on traditional health remedies, conservation and sustainable use of biodiversity, cultivation, value addition, and development of standards for indigenous drugs. Unfortunately it is difficult to find comprehensive information in this sector at a global level **[14].** Hawkins **[1]** provided sources of medicinal plants of conservation concern worldwide.

**2.2 Related Works**

Efforts have been and still being made by various research institutions, botanic gardens in published literature and databases (including **[1], [13], [18] – [34]**)**,** and herbarium to document and conserve medicinal plants.

The New York Botanical Garden currently grows ten species of plants on the Federal Endangered Species List. They are striving to preserve rare and endangered plants and participate with other institutions in doing this. The National Collection of Endangered Plants contains seeds, cuttings, and whole plants of 496 rare plant species native to the United States. The collection is stored at 25 gardens and arboreta that form part of the Center for Plant Conservation (CPC). The Royal Botanic Gardens at Kew, United Kingdom, support six *ex situ* and *in situ* conservation projects. The activities range from acting as the U.K. scientific Authority for Plants for Convention on International Trade in Endangered Species of Wild Fauna and Flora (CITES), cooperating in the recovery and reintroduction of endangered species, and aiding in the production of management plans for sustainable development and protected areas. The Wrigley Memorial and Botanical Gardens at Catalina Island, California, is still another example. The Gardens' emphasis is on California island endemic plants. Many of these plants are extremely rare, with some listed on the Endangered Species List **[12]**.

MEDPHYT **[35]** was built to collect data on the complete European pharmaceutical and toxicological plant world whose representatives were determined by medical and therapeutic benefit. The focus of the database content was the plant with description of their botanical characteristics, and history of discovery of therapeutic use, etymology, and synonyms. Besides the botanical characterisation there was information on both medical relevant biochemical compounds and their physicochemical characteristics, and toxicological as well as pharmaceutical facts. Its future outlook included comprehensive 3D visualisation techniques, implementation of the system for mobile systems and expansion of the content. This would require some additional work on both new scientific approaches to the action of mode of the drugs on molecular level and the database system MEDPHYT. Addition of further tables for animals, bacteria and fungi would also be required.

None of these attempts incorporated text-to-speech, voice/video, and multilingual features to address different cultures, languages, and audiences comprehensively on a multimedia platform. This gap limits knowledge discovery on useful medicinal plants from different areas of origin to be propagated elsewhere in addressing common diseases and ailments. Promotion of scientific research towards obtaining clues and discovery of potential lead compounds and novel therapeutics from medicinal plants has consequently been retarded.

### 3. Materials and Methods
#### 3.1 Data Collection

The criteria used for our data collection were centered on multipurpose uses in primary health care, needs for cultivation packages, potential in processing at the primary level, knowledge of availability and threat to the wild resources, and the need for collaborative work for understanding the patterns of diversity in relation to medicinal value and use.

The ethnobotanical data on plant species, families, vernacular names, prevalence, sustainable availability, plant parts used, medicinal usage by local communities and modalities of use, conditions for cultivation, phytochemical and pharmacological properties were collected, recorded, and discussed.

Medicinal plant materials were obtained randomly through personal contacts in the field, forestry-and-plant-science-based research institutions, local markets (*elewe-omo*), and at the homes of complementary and alternative medical practitioners (*babalawos*) in January - March 2010, June – September 2010, and December 2010 - February 2011 in Ota and Abeokuta towns in Ogun State of Nigeria.

Medicinal plant uses were discussed in detail with informants, after seeking prior informed consent from each respondent. Two hundred and fifty complementary and alternative medicine practitioners who know and use medicinal plants for treating various diseases were interviewed.

Following a semi-structured interview technique randomly administered, respondents were asked to provide detailed information about the vernacular plant name in Yoruba language; plant properties; harvesting region; ailments for which a plant was used; best harvesting time and season; plant parts used, as well as mode of preparation and application.

Older individuals, local medicine men or herbalists and others who claim to have effective prescriptions were interviewed. All interviews were carried out in Yoruba language, with at least one of the authors present. Three of the authors were fluent in Yoruba language, and no interpreter was needed to conduct the interviews.

**Table 5** shows some of the medicinal plants' characteristics gathered, while **Table 6** presents samples of the collected medicinal plants.

**Table 5 Data Collection Parameters (Extracts)**

| Scientific/BotanicalName | Family Name | CommonName | Synonyms | Local Names(Yoruba Lang) | Description | Medicinal Uses | Parts Used | Area(s) of Origin | Preparations / Dosage | Contraindications | Phytoconstituents | Adverse Reactions | Toxicity | Pharmacology | Drug interactions | Picture | Published Source(s) |
|---|---|---|---|---|---|---|---|---|---|---|---|---|---|---|---|---|---|
| | | | | | | | | | | | | | | | | | |

*Extinct, Vulnerable, Threatened, Endangered, Available, Rare

**Table 6 Extracts from the collected/analysed medicinal plants**

| Species | Family | Yoruba Lang | Parts Used | Use | Photo |
|---|---|---|---|---|---|
| ACA | facx | Jinwini | root decoction | WI | 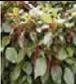 |
| AGR | fagx | Imi-esu, Akayunyun | leaf decoction | URT WI | 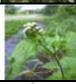 |
| ALL | falx | Alubosa ayu | Root | STR EYE | 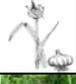 |
| ASP | | Aluki, Eye-kosun-Dangi | Root | MI | 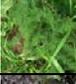 |
| ELY | fact | Ewe-Eso | Whole plant. | GNO IMP INF | 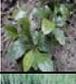 |
| EUP | feux | Orowere, Enuopire Enukopure | Leaves, exudate | DMT INF | 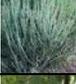 |
| FIC | fmox | Opoto, Farin bauree, Anwerenwa | Leaves stem, root, fruits | OED LEP EPL RIC INF | 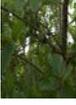 |
| ZNG | fznx | Jinja, Atale, Atalekopa | Rhizome Root | AST PIL, HEP OBE ANN CAN DYS | 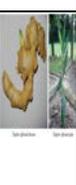 |

**Legend**: ANA = Anaemia, AST = Asthma, CAN = Cancer,

DMT = Dermatosis, DYS = Dysmenorrhoea, EPL = Epilepsy, EYE = Eye pain, GNO = Gonorrhoea, HEP = Hepatitis, IMP = Impotence, INF = Infertility, PIL = Piles, LEP = Leprosy, MI = Male Infertility, OBE = Obesity, OED = Oedema, RIC = Rickets, STR = Stroke, URT = Urinary Tract Infetcions, *WI* = Women Infertility
ACC = *Acalypha villicaulis* Hoschst, *AGR* = *Ageratum conyzoides L,* ALL = *Allium sativum* L., ASP = *Asparagus racemosa,* ELY = *Elytraria marginata,* EUP = *Euphorbia laterifolia,* FIC = *Ficus capensis* Thunb , ZNG = *Zingiber officinale* Rosc
facx = *Euphorbiaceae,* fagx = *Asteraceae,* falx = Alliaceae, fact = *Acanthaceae,* feux = *Euphorbiaceae,* fmox = *Moraceae,* fznx = Zingiberaceae

Additional data on medicinal plants were also collected from published literature: **[18] - [25]**; phytochemical databases; and reputably cognate internet-based medicinal plants' sources including **[26]-[34]** and shown in **Fig. 3**.

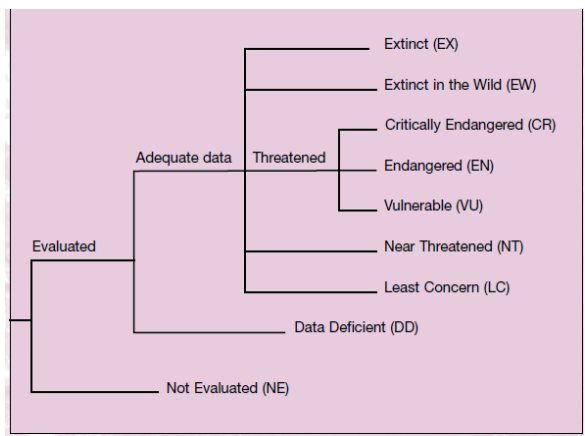

Fig. 4 The 2007 IUCN Red List of threatened species categories [1]

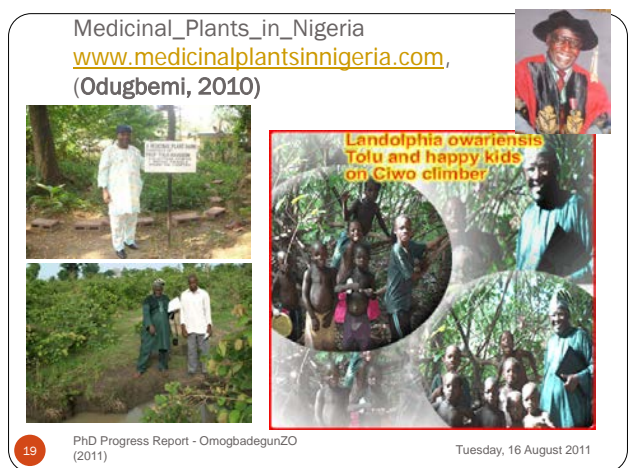

**Fig. 3 Medicinal Plants in Nigeria [30]**

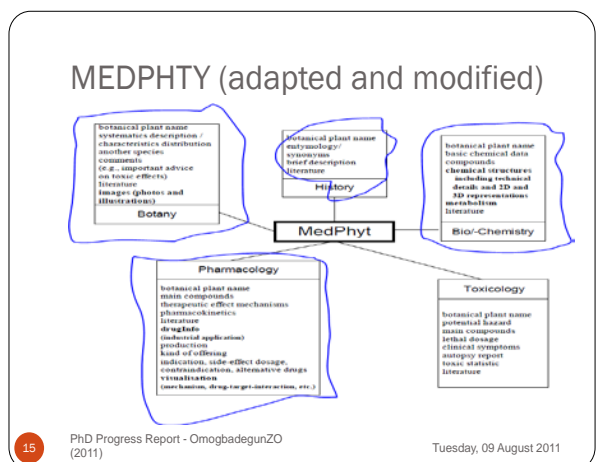

**Fig. 5 MEDPHTY adapted [35]**

### 3.2 Software Development

Guided by Threatened Species Categorization Standards in **Fig. 4 [1]**, adapting and modifying MEDPHYT of **Fig. 5 [35]** and DeeprootPlantBase of **Fig. 6** prototypes**,** the following were used: Visual Studio.Net and C# 3.0 Programming Language for creating the application, Microsoft Access for creating and querying the Database, HTML for displaying the plant Information in a text format, and Microsoft Jet Engine 4.0 to connect the application to the Database. Loquendo Multilingual Text-to-Speech Software was incorporated for audio content of the plant's salient characteristics, while YouTube and VLC Media Player were used for showing and playing the video of each plant chosen.

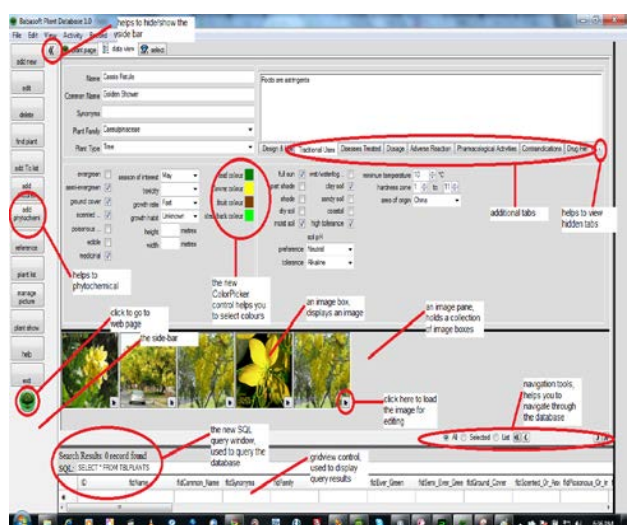

**Fig. 6 Enhanced Medicinal Plants Database after adaptation from Deeproot Plantbase**

## 4. Results

Over 250 medicinal plants were collected, analysed, and discussed. The entire plants with their flowers, fruits, and roots were collected and photo/snap taken. All the collected plants' details were used to populate the database of the developed system to address the medicinal plants' extinction problem as reflected in **Figs 6 - 7.**

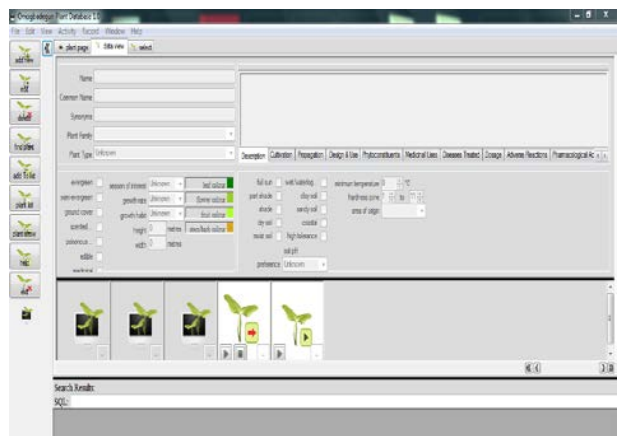

**Fig. 7** Multimedia-based Medicinal Plants Sustainability Management System (ONI_MMPSMS)

Details of each plant included scientific name, common name(s), family, multiple areas of origin, phytoconstituents, traditional medicinal uses, diseases treated, pharmacological activities, prescriptions, dosage, mode of preparation and administration, adverse reactions, toxicity, contraindications, drug-herb interactions, graphics, text-to-speech of contents, images, phytochemical structure, etc. A typical example extracted from **Table 6** and **Fig. 8**, *Zingiber officinale* Rosc (Common Ginger, - *Jinja, Atale, Atalekopa),* reflects Ginger (Zingiber Officinale, *Zingiberaceae* family) as originating from southern Asia. Nowadays, it is cultivated and commercialized around the world, particularly in China, India, Indonesia and Africa to demonstrate multiple areas of availability.

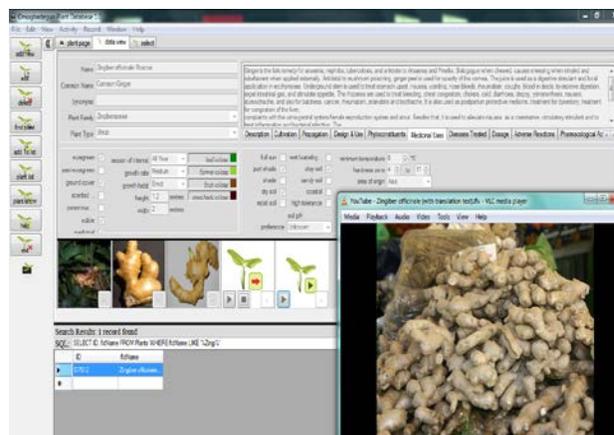

**Fig. 8** Enquiry from ONI_MMPSMS showing Diseases treated with a plant

The three major producing countries are India, China, and Nigeria. Technological and socioeconomic factors for cultivating *Zingiber officinale* Rosc were captured and stored. Among others, it was evident that more than one medicinal plant could be used in treating the same disease/ailment. For example, in **Table 6**, *Acalypha villicaulis Hoschst* and *Ageratum conyzoides* which differ only in the parts used treat women infertility. Similarly, *Elytraria marginata* and *Euphorbia laterifolia* handle infertility in both male and female.

In a similar vein, a single medicinal plant has multi-purpose use in handling more than one disease/ailment. Examples from **Table 6** showed *Elytraria marginata (for* Gonorrhoea, impotence, infertility*), Euphorbia laterifolia (for* Dermatosis, infertility*), Ficus capensis* Thunb *(for* Dysentery, oedema, leprosy, epilepsy, rickets, infertility), *Zingiber officinale* Rosc (for asthma, stimulant, piles, hepatitis, liver diseases, obesity, typhoid, anaemia, cancer, dysentery, Dysmenorrhoea).

Comprehensively on medicinal uses as documented in the database, Ginger is the folk remedy for anaemia, nephritis, tuberculosis, and antidote to Arisaema and Pinellia. Sialogogue when chewed, causes sneezing when inhaled and rubefacient when applied externally. Antidotal to mushroom poisoning, ginger peel is used for opacity of the cornea. The juice is used as a digestive stimulant and local application in ecchymoses. Underground stem is used to treat stomach upset, nausea, vomiting, nose bleeds, rheumatism, coughs, blood in stools, to improve digestion, expel intestinal gas, and stimulate appetite. The rhizomes are used to treat bleeding, chest congestion, cholera, cold, diarrhoea, dropsy, dysmenorrhoea, nausea, stomachache, and also for baldness, cancer, rheumatism, snakebite and toothache. It is also used as postpartum protective medicine, treatment for dysentery, treatment for congestion of the liver, complaints with the urino-genital system/female reproduction system and sinus. Besides that, it is used to alleviate nausea, as a carminative, circulatory stimulant and to treat inflammation and bacterial infection. The Commision E approved the internal use of ginger for dyspepsia and prevention of motion sickness. The British Herbal Compendium indicates ginger for atonic dyspepsia, colic, vomiting of pregnancy, anorexia, bronchitis and rheumatic complaints. European Scientific Cooperative on Phytotherapy (ESCOP) indicates its use for prophylaxis of the nausea and vomiting of motion sickness and to alleviate nausea after minor surgical procedures.

From this work, on **contraindications** for *Zingiber officinale* Rosc, it has been documented that users should consult physician before using ginger preparations in patients with blood coagulation disorders, taking anticoagulant drugs or with gallstones.

With respect to **toxicity**, ginger is a safe drug without any adverse reactions and has a wide range of utility. However, dried rhizomes during pregnancy should be avoided.

Fresh and dry ginger are tolerant and could be used as such. Generally, ginger is not subjected to any **purification** methods. Methods of purification for dry ginger and fresh juice are available from *Arogyakalpadruma* (an Ayurvedic text that concentrates on pediatrics). Purification of ginger may therefore be intended only for pediatric use, that is, to reduce the potency and pungency for infant use **[36].**

With respect to **drug-herb interactions**, it has documented that *Zingiber officinale* Rosc interacts with anticoagulants such as heparin, warfarin, drugs used in chemotherapy and ticlopidine. Ginger taken prior to 8-MOP (treatment for patients undergoing photopheresis) may substantially reduce nausea caused by 8-MOP. Ginger appears to increase the risk of bleeding in patients taking warfarin. However, ginger at recommended doses does not significantly affect clotting status, the pharmacokinetics or pharmacodynamics of warfarin in healthy subjects. Ginger also significantly decreased the oral bioavailability of cyclosporine. All the above characteristics have been captured for each medicinal plant as exemplified in **Fig. 8.**

The novelty in this work included multilingual text-to-speech (voice) and video features in a collaborative virtual environment for seamless exchange of information among scholars and practitioners of complementary and alternative medicine.

In addition, users of the system could search for any plant by any combination of its characteristics. The matched items are displayed in a separate window for the user to pick the desired plant. Furthermore, an interface exists for a sophisticated computer database user to use SQL SELECT statement in his/her search exercise. Plant search by other parameters such as area of origin is featured. Standard update actions in software engineering are provided. The video of the salient details of the retrieved plant could be viewed, and the audio feature could also be played as shown in **Figs 8 - 9**.

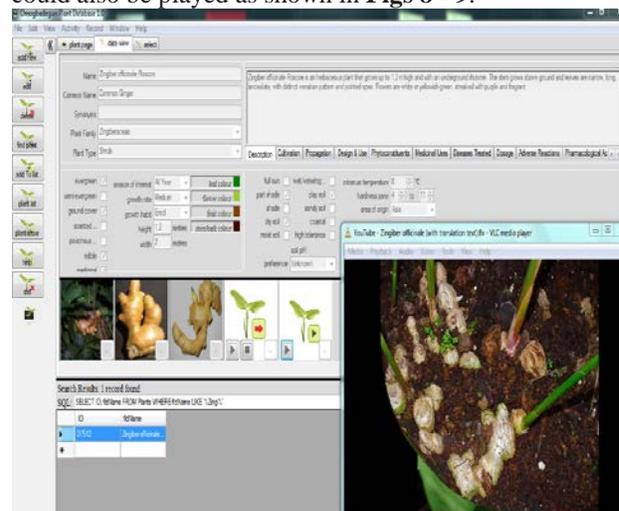

Fig. 9 Screen shot of ONI_MMPSMS features

Specific database of medicinal plants for each disease could be extracted and expanded from the large system.

5. **Discussion**

Conservation of threatened medicinal plants has become an increasingly important role. Medicinal plants' value that complements orthodox medicine in affordable manner has been documented for sustainability. The information made available on multiple areas of origin for cultivation elsewhere in solving medicinal plants' extinction challenge would also promote scientific research towards obtaining clues and discovery of potential lead compounds and novel therapeutics.

Beneficiaries of the software would include computer scientists, information technologists, biochemists, biologists, chemists, ethnobotanists, phytochemists, phytopharmacists, physicians and African traditional doctors (typically herbalists) to exchange information in African healing process. New plants would be added seamlessly as characteristics are assembled.

Among other characteristics, the following medicinal uses of *Zingiber officinale* Rosc published by **[37]** have been documented in the database as shown in **Fig. 8**: "Ginger (rhizome of *Zingiber officinale*) is a well known herb for its culinary and wide range of medicinal uses and is considered an essential component of the kitchen pharmacy. More commonly, ginger has been traditionally used in disorders of the gastrointestinal tract, as a stomachic, laxative, sialogogue, gastric emptying enhancer, appetizer, antiemetic, antidyspepsic, antispasmodic, and antiulcer agent with sufficient scientific support. Similarly, ginger has been shown to exhibit anti-inflammatory, hypoglycemic, antimigraine, antioxidant, hepatoprotective, diuretic, hypocholesterolemic, and antihypertensive activities.

Recently, ginger has gained wide attention for its therapeutic role as a safe and effective preventive treatment option for nausea and vomiting of pregnancy. Ginger has a long history of safety, as it has been used for centuries not only for medicinal purposes, but also as a food and spice. Although some health risk and safety concerns exist in the literature about its use by pregnant women, the clinical evidence of harm is lacking. Ginger might, therefore, be used as an effective treatment option for nausea and vomiting during pregnancy" **[37].** It could be deduced that *Zingiber officinale* Rosc could serve as a single medicine for internal use, as an ingredient in compound medicines, for external use, as an adjuvant, as an antidote, and for the purification of some mineral drugs.

## 6. Conclusions

The emergence of new infectious, chronic and drug-resistant diseases have prompted scientists to look towards medicinal plants as agents for treatment and prevention to foster high-quality and high-efficiency primary care. A system that documents and maintains comprehensive details on medicinal plants has been discussed. This work has provided the general public valuable insights into availability of and the traditional use of plants as medicines. This research effort would no doubt lead to a greater sense of confidence in many of the leading botanical raw materials of African origin in the medicinal plant trade. With comprehensive information provided on medicinal plants, this system would serve as a vehicle for critical gap-filling in research, study and application of research results in medicinal plants. Finally, the work provides a database for full scientific research towards obtaining clues and discovery of novel therapeutics.

### 6.1 Recommendations

In view of the multi-purpose use of medicinal plants in enhancing health, the following strategies to bridge the gap in the sustainability of ginger advanced by **[37]** are highly recommended for all medicinal plants:

• Enhancement of potential and realizable productivity through an integrated system of cultivation using high-yielding and resistant varieties, plant nutrient management, production technology suited to different agroecological situations and cropping systems, and need-based plant protection measures are future areas of increasing relevance in boosting ginger production.
• Resistance breeding against devastating diseases such as soft rot and bacterial wilt incorporating genes from wild relatives using biotechnological tools or through the exploitation of somaclonal variations.
• Enhancement of quality and evolving and popularizing very efficient postharvest handling techniques including product diversification.
• Tailoring production to meet export needs and international requirements such as clean ginger.
• Organic ginger production is to be promoted as a large quantity of immature ginger and fresh ginger are being used for the production of fresh ginger products. Popularization of products such as ginger beer, ginger ale, ginger squash, and ginger tea among consumers can help to create demand for ginger, which in turn will boost the ginger economy.

### 6.2 Future Development

A mobile version of the system deployable on a 4G network technology is underway. This would be accessible by complementary and alternative medicine (CAM) practitioners in the rural and underserved population areas. The software would also be made available in Nigeria's principal local languages (Hausa, Igbo, and Yoruba) and French in no distant future.

### Acknowledgment
Z.O. Omogbadegun, C.O. Uwadia, C.K. Ayo, V.W.Mbarika, N.Omoregbe, E.Otofia, and F. Chieze appreciate International Institute of Tropical Agriculture (IITA), Ibadan, Nigeria; Forestry Research Institute (FRI), Ibadan, Nigeria; Obafemi Awolowo University Herbarium, Ile-Ife, Nigeria; Dr. Abiodun Adebayo (Department of Biological Sciences, Covenant University, Ota, Nigeria), Geoff Looker (Deeproot Software Ltd, London, UK), Prof. Jiang Ding-Bi (Institute of Information on Traditional Chinese Medicine, China Academy of Chinese Medical Sciences, Beijing, P.R.China), and Prof. Yulan He (Informatics Research Centre, University of Reading, Reading RG6 6BX, UK).

### References
[1] A. Hawkins, *Plants for life*: *Medicinal plant conservation and botanic gardens,* Richmond, U.K Botanic Gardens Conservation International, 2008
[2] B. Kasirajan; R. Maruthamuthu; V. Gopalakrishnan; K. Arumugam; H. Asirvatham; V. Murali; R. Mohandass; and A. Bhaskar, "A database for medicinal plants used in treatment of asthma", *Bioinformation* 2(3): 2007, pp. 105-106.
[3] F. Cho-Ngwa, M. Abongwa, M. Ngemenya, and K.D. Nyongbela. "Selective activity of extracts of *Margaritaria discoidea* and *Homalium africanum* on *Onchocerca ochengi*", *BMC Complementary and Alternative Medicine* 2010, 10:62doi:10.1186/1472-6882-10-62
[4] P.A. Babu; G. Suneetha; R. Boddepalli; V.V. Lakshmi; T.S. Rani; Y. RamBabu; and K. Srinivas. "A database of 389 medicinal plants for diabetes", *Bioinformation* 1(4):, 2006, pp. 130-131
[5] D.N.Tewari, Report of the Task Force on Conservation & Sustainable use of Medicinal Plants, Government of India, Planning Commission, March – 2000, http://planningcommission.nic.in/aboutus/taskforce/tsk_medi.pdf Accessed February 7, 2011.

[6] T. E. Mafimisebi and A. E. Oguntade. "Preparation and use of plant medicines for farmers' health in Southwest Nigeria: sociocultural, magico-religious and economic aspects", *Journal of Ethnobiology and Ethnomedicine*, 6:1, 2010.

[7] A. P. Ekanem and F. V. Udoh. The Diversity of Medicinal Plants in Nigeria: An Overview. In Chi-Tang Ho (Ed). *African Natural Plant Products: New Discoveries and Challenges In Chemistry and Quality (ACS Symposium Series),* Oxford University Press, USA, 2010

[8] S. B. Kayne, *Complementary and Alternative Medicine*, Second edition, London: Pharmaceutical Press, 2009.

[9] E. Oladipo, M. G. Ogbe, Norman Molta, David Ladipo, Gamaniel Shingu, et al. Current *Status* of Biodiversity in *Nigeria:* Nigeria First National Biodiversity Report, 2001. www.cbd.int/doc/world/ng/ng-nr-01-en.doc. Accessed March 12, 2011.

[10] S. Agrawal and A. Chakrabarti. Potential Nutraceutical Ingredients from Plant Origin. In Yashwant Pathak *(Ed). Handbook of Nutraceuticals Volume 1: Ingredients, Formulations, and Applications, CRC Press, Taylor* & Francis Group, *2010.*

[11] V. Brower, "Back to Nature: Extinction of Medicinal Plants Threatens Drug Discovery", *JNCI Journal of the National Cancer Institute*, 2008, 100 (12), pp. 838-839

[12] M.J. Bogenschutz-Godwin, J.A. Duke, M. McKenzie, and P.B. Kaufman. *Plant Conservation.* In L.J. Cseke, A. Kirakosyan, P.B. Kaufman, S.L. Warber, J. A. Duke and H. L. Brielmann (Eds). *Natural Products from Plants* Second Edition, Taylor & Francis Group, 2006, pp 503-534

[13] M. Hamayun, S.A. Khan, E.Y. Sohn, and In-Jung Lee. "Folk medicinal knowledge and conservation status of some economically valued medicinal plants of District Swat, Pakistan", L*yonia, a journal of ecology and application*, 2006, 11(2), pp 101-113

[14] K. Vasisht and V. Kumar. *Compendium of Medicinal and Aromatic Plants AFRICA*, United Nations Industrial Development Organization and the International Centre for Science and High Technology, 2004

[15] V. A. J. Adekunle. "Conservation of Tree Species Diversity in Tropical Rainforest Ecosystem of South-West Nigeria", *Journal of Tropical Forest Science,* 2006, 18(2): pp. 91–101

[16] G. J. Osemeobo. "Can the Rain Forests of Nigeria Sustain Trade in Medicinal Plants?", *International Journal of Social Forestry,* 2010, Volume 3, Number 1, pp. 66-80.

[17] www.endangeredspecie.com/Why_Save_.htm. Accessed January 25, 2011

[18] O.O. G Amusan, Herbal Medicine in Swaziland: An Overview. In Chi-Tang Ho (Ed) *African Natural Plant Products: New Discoveries and Challenges In Chemistry and Quality (American Chemical Society Symposium Series)*, Oxford University Press, USA, 2010, pp. 26-44

[19] M.S. Dama, S.P. Akhand, and S. Rajender. "Nature versus nurture – plant resources in management of male infertility", *Frontiers in Bioscience* E2, 2010, 1001-1014, 1001, From Endocrinology Division, Central Drug Research Institute, Council of Scientific and Industrial Research, Lucknow, India - 226001

[20] T. L. Dog, Chaste Tree Extract in Women's Health: A Critical Review. In R. Cooper and F. Kronenberg (Eds). *Botanical Medicine: From Bench To Bedside,* Mary Ann Liebert, Inc 2010

[21] A.J. Hywood. Fertility Challenges. In A. J. Romm (Ed). *Botanical Medicine for Women.*, Churchill Livingstone, an imprint of Elsevier Inc, 2010, Pp. 345-357

[22] M. Idu, J.O. Erhabor, and H.M. Efijuemue. "Documentation on Medicinal Plants Sold in Markets in Abeokuta, Nigeria", *Tropical Journal of Pharmaceutical Research,* April 2010; 9 (2), pp. 110-118

[23] J. D. Olowokudejo, A. B. Kadiri, and V.A. Travih. "An Ethnobotanical Survey of Herbal Markets and Medicinal Plants in Lagos State of Nigeria", *Ethnobotanical Leaflets, 2008, 12, pp. 851-865.*

[24] J. K. Rao, J. Suneetha, T.V.V. S. Reddi, and O. A. Kumar. "Ethnomedicine of the Gadabas, a primitive tribe of Visakhapatnam district, Andhra Pradesh", *International Multidisciplinary Research Journal* 2011, 1/2, pp.10-14

[25] World Health Organization*, WHO monographs on selected medicinal plants* volumes 1- 4, WHO Press, World Health Organization, 20 Avenue Appia, 1211 Geneva 27, Switzerland, 2009

[26] Natural Medicines Comprehensive Database, www.naturaldatabase.com

[27] Natural Standard, www.naturalstandard.com

[28] American Botanical Council, www.herbalgram.org

[29] Botanic Garden Conservation International, www.bgci.org

[30] T. Odugbemi, www.medicinalplantsinnigeria.com

[31] ScienceDirect, www.sciencedirect.com

[32] Biomed Central, www.biomedcentral.com

[33] Springerlink, www.springerlink.com

[34] PubMed, www.ncbi.nlm.nih.gov/entrez/query.fcgi?db=PubMed

[35] C. Kettner; H. Kosch; M. Lang; J. Lachner; D. Oborny; and E. Teppan. "Creating a Medicinal Plant Database", http://www.beilstein-institut.de/englisch/1024.htm. Accessed January 10, 2011

[36] P.N. Ravindran and K. N. Babu (Eds). *Ginger: the genus Zingiber*, CRC Press, 2005

[37] A.Ali and A.H. Gilani. "Medicinal Value of Ginger With Focus on Its Use in Nausea and Vomiting of Pregnancy", *International Journal of Food Properties*, 2007,10, pp. 269–278

Zacchaeus Oni **Omogbadegun** holds B.Sc (Hons) Computer Science (Second Class Upper Div, 1979) from University of Ibadan, Ibadan, Nigeria and M.Sc Computer Science (2003) from University of Lagos, Lagos, Nigeria. He has over thirty years of progressively cognate experience in Information Technology spanning Software Engineering, Education and Training. He has worked either as an employee or Information Technology Consultant in bluechip industrial organizations within and outside Nigeria: Mobil Oil Nigeria Ltd (1980-1990), Equitorial Trust Bank (1990-1994), and as an Information Technology Consultant to: Ghana's Social Security & National Insurance Trust Software Replacement, Accra, Ghana, (1994-1996); Information Technology Consultant [Contract], FAO of the United Nations, Regional Office for Africa, Accra, Ghana ( 06/1997 – 09/1997); Edo State Government of Nigeria's Ministry of Education (01/1998- 12/1999); Ondo State Government of Nigeria & UNDP (Sept; 2000); Ekiti State Government of Nigeria's Ministry of Health (Sept; 2001). He has represented his employers in Tanzania, United Kingdom, and Zimbabwe at Software Internationalization Project programmes (1985-1997). He joined academics as Lecturer I (Computer Science) at The Federal

Polytechnic, Ado-Ekiti, Ekiti State, Nigeria (June 2001 – September 2005). He joined Covenant University, Ota, Ogun State, Nigeria as Lecturer II (October 2005) and became Lecturer I (2008 - present). He is currently a PhD (Computer Science) student in the Department of Computer and Information Sciences, Covenant University, Ota, Ogun State, Nigeria. His research interests are Software Engineering (Formal Methods), Healthcare Informatics, MDG's Education and Health-related goals, Artificial Intelligence, Education Informatics, Neural Networks, Biomedical Engineering, Computer Security, Multimedia Database System, and Information Technology Laws. His presentations at both local and international conferences included "Security in Healthcare Information Systems", "Impact of Mobile and Wireless Technologies on Healthcare Delivery Services", and "3G and 4G Technologies' Framework for Realising Millenium Development Goals in Healthcare in Nigeria", He is a member of the Nigerian Computer Society (NCS), and Computer Professional Registration Council of Nigeria (CPN).

Professor Charles Onuwa **Uwadia** holds a BSc. Degree honours in Computer Science from the University of Ibadan in 1979. He had his MSc. in 1983 and PhD in 1990 both at the University of Lagos. Professor Uwadia joined the services of the University of Lagos as an Assistant Lecturer in 1983, and rose steadily to the post of a full Professor of Computer Science in the year 2004. His major area of specialization is Software Engineering with emphasis on Compiling Techniques and Systems Software. He is also actively involved in teaching and research work in networking, congestion control and management aspects of Information Technology. With over 50 publications in both local and international journals, he has three books to his name.

He is an active member of the Nigeria Computer Society (NCS), and the Computer Professionals Registration Council of Nigeria. He is also a Fellow of NCS. He had served NCS in various capacities including: Chairman Education Committee (1997 – 1999); Chairman Conferences Committee (1999 – 2003); Chairman Publications Committee (2003 – July 2007). He is the current President of the Society. Besides his enviable service in Nigeria, Professor Uwadia has been a visiting Fellow to Universities and Institutions in the United States of America, Europe, Asia, and Australia. He has many publications to his credit. Professor Charles Onuwa Uwadia is the current Director of the Centre for Information Technology and Systems (CITS) of the University of Lagos; a position he has held since 2005.

Professor **Ayo** Charles Korede holds a B.Sc. M.Sc. and Ph.D in Computer Science. He is currently the Director, Academic Planning Unit of Covenant University. He was the pioneer Head of Computer and Information Sciences Department of the University. His research interests include: Mobile computing, Internet programming, eBusiness, eGovernment and Software Engineering. He is a member of the Nigerian Computer Society (NCS), and Computer Professionals (Registration Council) of Nigeria (CPN). Similarly, he is professionally certified in CISCO and Microsoft products. Prof. Ayo is a member of a number of international research bodies such as the Centre for Business Information, Organization and Process Management (BIOPoM), University of Westminster, London; the Review Committee of the European Conference on E-Government ECEG); the programme committee, IADIS Information Systems; the Editorial Board, Journal of Information and communication Technology for Human Development (IJICTHD), the Editorial Board, International Journal of Scientific Research in Education (IJSRE) and the Editorial Board, African Journal of Business Management amongst others. Furthermore, Prof. Ayo is an External Examiner to a number of Nigerian universities at both Undergraduate and Postgraduate levels in Ladoke Akintola University of Technology, Ogbomoso, the Redeemers University, Ogun State, Bells University of Technology, Ota, etc. He has supervised about 200 postgraduate projects at Postgraduate Diploma, Masters and Ph.D levels, and he has several publications in scholarly journals and conferences.

**Third Author** is a member of the IEEE and the IEEE Computer Society. Do not specify email address here.